\documentclass[12pt]{article}
\newcommand{\g}{\gamma}
\newcommand{\q}{\rho}
\newcommand{\n}{\nu}
\newcommand{\m}{\mu}
\newcommand{\al}{\alpha}
\newcommand{\be}{\beta}
\newcommand{\la}{\lambda}
\newcommand{\s}{\sigma}
\newcommand{\hh}{\tilde{\cal H}}

\begin{document}
\vspace*{-.6in}
\thispagestyle{empty}
\begin{flushright}
CALT-68-2093\\
hep-th/9701166
\end{flushright}
\baselineskip = 20pt

\vspace{.5in}
{\Large
\begin{center}
World-Volume Action of the M Theory Five-Brane\footnote{Work
supported in part by the U.S. Dept. of Energy under Grant No.
DE-FG03-92-ER40701.}
\end{center}}

\begin{center}
Mina Aganagic, Jaemo Park, Costin Popescu, and John H. Schwarz\\
\emph{California Institute of Technology, Pasadena, CA  91125, USA}
\end{center}
\vspace{1in}

\begin{center}
\textbf{Abstract}
\end{center}
\begin{quotation}
\noindent  This paper presents a 6d world-volume action that describes the
dynamics of the M theory five-brane in a flat 11d space-time background. The world-volume
action has global 11d super-Poincar\'e invariance, as well as 6d general coordinate invariance
and kappa symmetry, which are realized as local symmetries. The paper mostly
considers a formulation in which general coordinate invariance is not manifest in
one direction. However, it also describes briefly an alternative formulation,
due to Pasti, Sorokin, and Tonin, in which general coordinate invariance is
manifest. The latter approach requires auxiliary fields and new gauge invariances.

\end{quotation}
\vfil

\newpage

\pagenumbering{arabic}

\section{Introduction}

World-volume actions of $p$-branes encode much information about their dynamics.
In the case of strings (in flat backgrounds) the world-volume theory has been
quantized and used to construct the string perturbation expansion.  In the case
of $p$-branes with $p > 1$, one does not expect that it is possible to do the
same.  Still, many recent works have shown that an understanding of $p$-branes,
including their excitations, can be very useful.  Much non-perturbative
information has been gleaned by considering vacua containing various branes of
infinite extension. (A good example is provided by the 7-branes of F-theory~\cite{vafa}.)
Also, non-perturbative excitations described by wrapping
$p$-branes about various cycles have played a central role in recent studies of
black hole entropy as well as other problems~\cite{becker,strominger,harvey}.  
We suspect that a more detailed
characterization of $p$-brane world-volume dynamics will enable these studies to
go further.

The actions for the class of supersymmetric $p$-branes
whose only degrees of freedom are the
superspace coordinates $X$ and $\theta$ of the ambient space-time were
constructed during the decade of the 1980's~\cite{green1,hughes,bergshoeff2,achucarro1,duff2}.  
Much more recently, the actions
for D-branes in type II theories have been 
constructed~\cite{cederwall1,aganagic,cederwall2,bergshoeff1}.  
In addition to the $X$ and 
$\theta$ variables, these world-volume theories contain a U(1) gauge field 
with Born--Infeld self interactions~\cite{born,fradkin,bergshoeff3,abouelsaood,leigh,bachas}.  
For maximally supersymmetric
theories, the only significant $p$-brane action that remains to be formulated is
that of the M theory five-brane~\cite{guven,bergshoeff6,witten2,berkooz,howe}.  
This paper presents the solution.

The new feature that makes the M theory five-brane example somewhat more
challenging than the other ones is the presence of a second-rank tensor gauge
field, in addition to the $X$ and $\theta$ coordinates~\cite{callan}.  This gauge field
describes a chiral boson in the world volume, since its field strength is
self-dual in the linearized approximation.  It has been known for a long time
that there is no straightforward way to construct a covariant action that
describes propagation of the self-dual part of this field without also bringing
in the anti-self-dual part~\cite{marcus}.  Various proposals for dealing with this problem
have been suggested over the years.  The main one that we adopt is based on a
formulation in which general coordinate invariance is only manifest in five of
the six dimensions~\cite{henneaux,schwarz2,verlinde,perry}.  
It is also present in the sixth direction, but the
transformation formulas that describe the symmetry are rather complicated.  The
bosonic part of the five-brane theory, constructed by this method, has been
presented recently~\cite{jhs}.  Another approach to the problem of the chiral boson uses
an infinite number of auxiliary fields~\cite{mcclain,berkovits,bengtsson}.

Very recently, a manifestly covariant formulation involving only a finite
number of auxiliary fields (and compensating gauge invariances) has been
introduced by Pasti, Sorokin, and Tonin~\cite{pasti1,pasti2}.  
Constructions using the PST
formulation  turn out to be about as complicated as those in the formulation
without manifest covariance.  In fact, one of the new gauge invariances of the
PST formulation involves the same subtleties as those of general coordinate
invariance in the non-covariant approach, since one can gauge fix the PST
formulas to obtain the non-covariant ones and show that compensating gauge
transformations are the origin of the complicated general coordinate
transformation.

Besides general coordinate invariance, the other essential symmetry of the
world-volume theory of any super $p$-brane is a fermionic symmetry called kappa
symmetry.  It is always needed to remove half the degrees of freedom carried
by the $\theta$ variables, leaving altogether eight propagating fermionic
degrees of freedom.  This is the same as the number of bosonic degrees of
freedom, of course, as required by supersymmetry.  The way this is achieved is
by adding a suitable Wess--Zumino term to the action.

In all previous super $p$-brane examples, the global super-Poincar\'e symmetry
(induced from an ambient flat space-time background) is implemented separately
for the Wess--Zumino term and the other terms.  The story in the case of the
M theory five-brane has a surprising new feature.  Namely, extending the
bosonic five-brane theory to achieve global 11d super-Poincar\'e symmetry
uniquely determines the complete action, including the Wess--Zumino term.  The
formula obtained in this way is then shown to have general coordinate
invariance and local kappa symmetry.  
In the covariant PST formulation one is forced to organize the
terms somewhat differently, so in that approach the story looks somewhat more
conventional.  Specifically, the covariant action divides naturally into two
pieces: one piece is the supersymmetrized bosonic theory and the second is a
separately supersymmetric Wess--Zumino term.  The reason 
these statements are not in
contradiction is that the PST gauge invariances, which are needed to
achieve the right {\it bosonic} degrees of freedom, require that both terms be
included.

This paper is organized as follows.  Section 2 reviews the construction of the
bosonic part of the M theory five-brane action in both the non-covariant and
the PST formulations.  Section 3 then describes the supersymmetrization of this
theory and the determination of the Wess--Zumino term in the non-covariant
formulation.  The proof that the resulting theory has (non-manifest) general
coordinate invariance is given in Section 4.  Section 5 presents the proof of
kappa symmetry.  The verification of two crucial identities is relegated to a
pair of appendices.  This section also sketches the corresponding formulas in the PST
formulation.  Section 6 describes double dimensional reduction, which gives
rise to a 4-brane in 10d space-time.  The resulting theory gives a dual
formulation of the D4-brane of type IIA theory in which the
theory is expressed in terms of a two-form gauge field instead of the dual U(1)
vector gauge field.  Some concluding remarks are made in Section 7.

\medskip
\section{Review of the Bosonic Theory}

\subsection{Formulation Without Manifest Covariance}

Ref. \cite{jhs} analyzed the problem of coupling a 6d self-dual tensor gauge field to a
metric field so as to achieve general coordinate invariance. 
It presented a formulation in which one direction is treated differently from
the other five. At the time that work was done,
the author knew of no straightforward way to make the general
covariance manifest. However, shortly thereafter a paper appeared~\cite{pasti1} that 
presents equivalent results using a manifestly covariant formulation~\cite{pasti2},
which we refer to as the PST formulation.
The relation between the two approaches will be described in the next subsection.
As one might expect, they entail similar complications and there does not
appear to be much advantage to one approach over the other. Therefore, we 
will present the supersymmetric M theory 5-brane action in the formulation
without manifest covariance. This action corresponds to a partially
gauge-fixed version of the corresponding action in the PST formulation.

In the present work we denote the 6d (world volume) coordinates by 
$\sigma^{\hat\mu} = (\sigma^\mu, \sigma^5)$,
where $\mu = 0,1,2,3,4$.  (In ref. \cite{jhs}
they were called $x^{\hat\mu}$.) The $\sigma^5$ direction is singled out as the one that
will be treated differently from the other five.\footnote{This is a
space-like direction, but one could also choose a time-like  one. (See
the discussion in sect. 2.2.)
The reason we prefer this choice is that in section 6, where we perform a double dimension
reduction to obtain a 4-brane in 10d, elimination of the special dimension leaves
manifestly covariant equations.}  The 6d metric
$G_{\hat\mu\hat\nu}$ contains 5d pieces $G_{\mu\nu}, G_{\mu 5}$, and $G_{55}$.
All formulas will be written with manifest 5d general coordinate invariance.
As in refs.~\cite{perry,jhs}, we represent the self-dual tensor gauge field by a
$5\times 5$ antisymmetric tensor $B_{\mu\nu}$, and its 5d curl by
$H_{\mu\nu\rho} = 3 \partial_{[\mu} B_{\nu\rho]}$. A useful quantity is the dual 
\begin{equation}
\tilde{H}^{\mu\nu} = {1\over 6} \epsilon^{\mu\nu\rho\lambda\sigma}
H_{\rho\lambda\sigma}.
\end{equation}

It was shown in ref.~\cite{jhs} that a class of generally covariant
bosonic theories could be represented in the form
$L = L_1 + L_2 + L_3$, where\footnote{The formula given in ref.~\cite{jhs}
has been rescaled by an overall factor of $-1/2$.}
\begin{eqnarray}
L_1 &=& -{1\over 2}\sqrt{-G} f(z_1,z_2), \nonumber \\
L_2 &=& -{1\over 4} \tilde{H}^{\mu\nu} \partial_5 B_{\mu\nu}, \\
L_3 &=&  {1\over 8}
\epsilon_{\mu\nu\rho\lambda\sigma} {G^{5\rho}\over G^{55}} \tilde{H}^{\mu\nu}
\tilde{H}^{\lambda\sigma}.\nonumber 
\end{eqnarray}
The notation is as follows:  $G$ is the 6d determinant $(G =
{\rm det}\, G_{\hat\mu\hat\nu})$ and
$G_5$ is the 5d determinant $(G_5 =
{\rm det}\, G_{\mu\nu})$, while $G^{55}$ and $G^{5\rho}$ are components of the inverse
6d metric $G^{\hat\mu \hat\nu}$.  The $\epsilon$ symbols are purely numerical with $\epsilon^{01234} = 1$ and
$\epsilon^{\mu\nu\rho\lambda\sigma} = - \epsilon_{\mu\nu\rho\lambda\sigma}$.  A
useful relation is $G_5 = G G^{55}$.
The $z$ variables are defined to be
\begin{eqnarray}
z_1 &=& {{\rm tr} (G\tilde{H} G\tilde{H})\over 2( -G_5)}\nonumber \\
z_2 &=& {{\rm tr} (G\tilde{H} G\tilde{H} G\tilde{H} G\tilde{H})\over 4 (-G_5)^2}.
\label{zdefs}
\end{eqnarray}
The trace only involves 5d indices:
\begin{equation}
{\rm tr} (G\tilde{H} G\tilde{H}) = G_{\mu\nu} \tilde{H}^{\nu\rho} G_{\rho\lambda}
\tilde{H}^{\lambda\mu}.
\end{equation}
The quantities $z_1$ and $z_2$
are scalars under 5d general coordinate
transformations.  

Infinitesimal parameters of general coordinate transformations are denoted
$\xi^{\hat\mu} = (\xi^\mu, \xi)$.  Since 5d general coordinate invariance is
manifest, we focus on the $\xi$ transformations only.  The metric transforms in
the standard way
\begin{equation}
\delta_\xi G_{\hat\mu \hat\nu} = \xi \partial_5 G_{\hat\mu \hat\nu} +
\partial_{\hat\mu} \xi G_{5\hat\nu} + \partial_{\hat\nu} \xi G_{\hat\mu 5}.
\label{Gvar}
\end{equation}
The variation of $B_{\mu\nu}$ is given by a more complicated rule, whose origin is
explained in ref.~\cite{jhs}:
\begin{equation}
\delta_\xi B_{\mu\nu} = \xi K_{\mu\nu}, \label{Bvar}
\end{equation}
where
\begin{equation}
K_{\mu\nu} = 2{\partial (L_1 + L_3) \over\partial \tilde{H}^{\mu\nu}} =
K_{\mu\nu}^{(1)} f_1+ K_{\mu\nu}^{(2)} f_2+ K_{\mu\nu}^{(\epsilon)}
\label{Kform1}
\end{equation}
with
\begin{eqnarray}
K_{\mu\nu}^{(1)} &=& {\sqrt{-G} \over (-G_5)}{(G\tilde{H} G)_{\mu\nu}}
\nonumber \\
K_{\mu\nu}^{(2)} &=& {\sqrt{-G} \over (-G_5)^2}{(G\tilde{H} G\tilde{H} G\tilde{H}
G)_{\mu\nu}}  \label{Kform2}\\
K_{\mu\nu}^{(\epsilon)} &=& \epsilon_{\mu\nu\rho\lambda\sigma}
{G^{5\rho}\over 2 G^{55}} \tilde{H}^{\lambda\sigma}, \nonumber
\end{eqnarray}
and we have defined
\begin{equation}
f_i = {\partial f\over\partial z_i} , \quad i = 1,2.
\end{equation}

Assembling the results given above, ref.~\cite{jhs} showed that
the required general coordinate transformation symmetry is
achieved if, and only if, the function $f$ satisfies the nonlinear partial
differential equation~\cite{gibbons}
\begin{equation}
f_1^2 + z_1 f_1 f_2 + \big({1\over 2} z_1^2 - z_2\big) f_2^2 = 1.
\end{equation}
As discussed in~\cite{perry},
this equation has many solutions, but the one of relevance to the
M theory five-brane is 
\begin{equation}
f = 2 \sqrt{1 + z_1 + {1\over 2} z_1^2 - z_2}.
\end{equation}
For this choice $L_1$
can reexpressed in the Born--Infeld form
\begin{equation}
L _1 = - \sqrt{- {\rm det} \Big(G_{\hat\mu \hat\nu} + i G_{\hat\mu\rho} G_{\hat\nu
\lambda} \tilde{H}^{\rho\lambda} / \sqrt{-G_5}\Big)} . \label{bosonicL1}
\end{equation}
This expression is real, despite the factor of $i$, because it is an even function of
$\tilde H$. Eliminating the factor of $i$ would correspond to replacing $z_1$
by $-z_1$, which also solves the differential equation. However, it is essential 
for the five-brane application that the phases be chosen as shown.

\subsection{The PST Formulation}

In ref.~\cite{pasti1} (using techniques developed
in ref.~\cite{pasti2}) results equivalent to those of the preceding subsection are
described in a manifestly covariant way.  To do this, the field $B_{\mu\nu}$ is
extended to $B_{\hat\mu \hat\nu}$ with field strength $H_{\hat\mu \hat\nu
\hat\rho}$.  In addition, an auxiliary scalar field $a$ is
introduced.  The PST formulation has new gauge symmetries (described below)
that allow one to choose the gauge $B_{\mu 5} = 0,$  $a = \sigma^5$
(and hence $\partial_{\hat\mu}a =
\delta_{\hat\mu}^5$).  In this gauge, the covariant PST formulas reduce to those
of sect. 2.1.

As will become clear, the scalar field $a$ is really a zero-form potential 
with one-form field strength
$da$. Only the field strength needs to be single-valued. Furthermore, for
the action to be nonsingular, it is necessary that the 6 manifold $M_6$ admit
nowhere null closed one-forms and that $da$ be restricted to the class of
such one-forms. It is allowed to be either time-like or space-like, however.
This topological restriction on $M_6$ is consistent with the conclusions reached
in ref.~\cite{witten2}

Equation (\ref{bosonicL1}) 
expressed $L_1$ in terms of the determinant of the $6 \times 6$ matrix
\begin{equation}
M_{\hat\mu\hat\nu} = G_{\hat\mu\hat\nu} + i {G_{\hat\mu \rho} G_{\hat\nu
\lambda}\over \sqrt{- GG^{55}}} \tilde{H}^{\rho\lambda}.
\end{equation}
In the PST approach this is extended to the manifestly covariant form
\begin{equation}
M_{\hat\mu\hat\nu}^{\rm cov.} = G_{\hat\mu\hat\nu} + i {G_{\hat\mu\hat\rho}
G_{\hat\nu \hat\lambda}\over\sqrt{-G (\partial a)^2}}
\tilde{H}_{\rm cov.}^{\hat\rho \hat\lambda}. \label{Mcov}
\end{equation}
The quantity
\begin{equation}
(\partial a)^2 = G^{\hat\mu\hat\nu} \partial_{\hat\mu} a \partial_{\hat\nu} a
\end{equation}
reduces to $G^{55}$ upon setting $\partial_{\hat\mu}a  = \delta_{\hat\mu}^5$,
and
\begin{equation}
\tilde{H}_{\rm cov.}^{\hat\rho\hat\lambda} \equiv {1\over 6} \epsilon^{\hat\rho
\hat\lambda \hat\mu \hat\nu \hat\sigma \hat\tau} H_{\hat\mu \hat\nu \hat\sigma}
\partial_{\hat\tau} a
\end{equation}
reduces to $\tilde{H}^{\rho\lambda}$.  Thus $M_{\hat\mu \hat\nu}^{\rm cov.}$
replaces $M_{\hat\mu\hat\nu}$ in $L_1$.  Furthermore, the expression
\begin{equation}
L' = - { 1\over 4(\partial a)^2} \tilde{H}_{\rm cov.}^{\hat\mu \hat\nu}
H_{\hat\mu\hat\nu\hat\rho} G^{\hat\rho\hat\lambda} \partial_{\hat\lambda} a,
\end{equation}
which transforms under general coordinate transformations as a scalar density,
reduces to $L_2 + L_3$ upon gauge fixing. It is interesting that $L_2$ and $L_3$ are
unified in this formulation.

Let us now describe the new gauge symmetries of ref.~\cite{pasti1}.  Since degrees of
freedom $a$ and $B_{\mu 5}$ have been added, corresponding gauge symmetries are
required.  One of them is
\begin{equation}
\delta B_{\hat\mu \hat\nu} = 2 \phi_{[\hat\mu} \partial_{\hat\nu]} a,
\end{equation}
where $\phi_{\hat\mu}$ are infinitesimal parameters, and the other fields do not
vary.  In terms of differential forms, this implies $\delta H = d\phi
da$.  $\tilde{H}_{\rm cov.}^{\hat\rho \hat\lambda}$ is invariant under this transformation,
since it corresponds to the dual of $H da$, but $da da = 0$.
Thus the covariant version of $L_1$ is invariant under this transformation.
The variation of $L'$, on the other hand, is a total derivative.

The second local symmetry involves an infinitesimal 
scalar parameter $\varphi$.  The transformation
rules are $\delta G_{\hat\mu\hat\nu} = 0, \delta a = \varphi$, and
\begin{equation}
\delta B_{\hat\mu\hat\nu} = {1\over (\partial a)^2} \varphi
H_{\hat\mu\hat\nu\hat\rho} G^{\hat\rho\hat\lambda} \partial_{\hat\lambda} a +
\varphi V_{\hat\mu\hat\nu},
\end{equation}
where the quantity $V_{\hat\mu\hat\nu}$ is to be determined.  This
transformation is just as complicated as the non-manifest general coordinate
transformation in the non-covariant formalism.  Rather than derive it from
scratch, let's see what is required to agree with the previous formulas after
gauge fixing.  In other words, we fix the gauge $\partial_{\hat\mu} a =
\delta_{\hat\mu}^5$ and $B_{\mu 5} = 0$, and figure out what the resulting
$\xi$ transformations are.  We need
\begin{equation}
\delta a = \varphi + \xi \partial_5 a = \varphi + \xi = 0,
\end{equation}
which tells us that $\varphi = - \xi$.  Then
\begin{eqnarray}
\delta_{\xi} B_{\mu\nu} &=& {1\over (\partial a)^2} \varphi H_{\mu\nu\hat\rho}
G^{\hat\rho\hat\lambda} \partial_{\hat\lambda} a + \varphi V_{\mu\nu} + \xi
H_{5\mu\nu}\nonumber \\
&=& - \xi \left({G^{\rho 5}\over G^{55}} H_{\mu\nu\rho} + V_{\mu\nu}\right) =
\xi (K_{\mu\nu}^{(\epsilon)} - V_{\mu\nu}).
\end{eqnarray}
Thus, comparing with eqs.~(\ref{Bvar}) and (\ref{Kform1}), we need the covariant definition
\begin{equation}
V_{\hat\mu\hat\nu} = - 2 {\partial L_1\over \partial
\tilde{H}_{\rm cov.}^{\hat\mu\hat\nu}}
\end{equation}
to achieve agreement with our previous results.

To summarize, we have learned that the covariant PST formulation has new gauge
transformations, and one of them encodes the complications that end up in
general coordinate invariance after gauge fixing.  Thus this formalism is not
simpler than the non-covariant one.  However, it is more symmetrical, and it
does raise new questions, such as whether there are other gauge choices that
are worth exploring.

\medskip

\section{Supersymmetrization}

The super--Poincar\'e symmetry of the flat 11d space-time background should be
implemented as a global symmetry of the five-brane theory.  In terms of
superspace coordinates $X^M$ and $\theta$, the 11d 
supersymmetry transformation is given by
\begin{equation}
\delta\theta = \epsilon \quad {\rm and} \quad
\delta X^M = \bar\epsilon \Gamma^M \theta.
\end{equation}
Our convention is that the index $M$ takes the values $M = 0, 1, \ldots, 9, 11$.
Skipping $M=10$ may seem a bit peculiar, but then $X^{11}$ is the 11th dimension.  Also,
the Dirac matrix $\Gamma_{11} = \Gamma_0 \Gamma_1 \ldots \Gamma_9$, which
appears in ten dimensions as a chirality operator, is precisely the matrix we
associate with the 11th dimension.  The spinors $\epsilon$ and $\theta$ are
32-component Majorana spinors.  The Dirac algebra is
\begin{equation}
\{\Gamma_M, \Gamma_N\} = 2\eta_{MN},
\end{equation}
where $\eta_{MN}$ is the 11d Lorentz metric with signature $(- ++ \ldots +)$.

As in other supersymmetric $p$-brane theories, two supersymmetric quantities are
$\partial_{\hat\mu} \theta$ and
\begin{equation}
\Pi_{\hat\mu}^M = \partial_{\hat\mu} X^M - \bar\theta \Gamma^M
\partial_{\hat\mu} \theta. \label{Pidef}
\end{equation}
The appropriate choice for the world-volume metric is then the supersymmetric
quantity
\begin{equation}
G_{\hat\mu \hat\nu} = \eta_{MN} \Pi_{\hat\mu}^M \Pi_{\hat\nu}^N.
\label{susymetric}
\end{equation}
Taking $\theta$ and $X^M$ to be scalars under world-volume general coordinate
transformations, $G_{\hat\mu \hat\nu}$ transforms in the standard way.

In addition, we require an appropriate supersymmetric extension of $H = d B$, which
we write as
\begin{equation}
{\cal H}_{\mu\nu\rho} = H_{\mu\nu\rho} - b_{\mu\nu\rho},
\end{equation}
or, in terms of differential forms, ${\cal H} = H - b_3$.  The idea is to
choose a $b_3$ whose supersymmetry variation is exact, so that it can be
cancelled by an appropriate variation of $B$.  The appropriate choice turns out to
be
\begin{equation}
b_3={1 \over 6} b_{\mu\nu\rho} d\sigma^{\mu} d\sigma^{\nu}d\sigma^{\rho}
=  {1\over 2} \bar\theta \Gamma_{MN} d\theta (dX^M d X^N + dX^M \bar\theta
\Gamma^N d\theta + {1\over 3} \bar\theta \Gamma^M d\theta \bar\theta \Gamma^N
d\theta). \label{b3form}
\end{equation}
Varying this, using $\delta_{\epsilon}\theta 
= \epsilon$ and $\delta_{\epsilon} X^M = \bar\epsilon
\Gamma^M \theta$, one finds that ${\cal H}$ is invariant for the choice
\[
\delta_\epsilon B = -{1\over 2} \bar\epsilon \Gamma_{MN} \theta (dX^M dX^N +
{2\over 3} \bar\theta \Gamma^M d\theta dX^N + {1\over 15} \bar\theta \Gamma^M
d\theta \bar\theta \Gamma^N d\theta)\]
\begin{equation}
- {1\over 6} \bar \epsilon \Gamma_M \theta \bar\theta \Gamma_{MN} d\theta (d
X^N + {1\over 5} \bar\theta \Gamma^N d\theta).
\end{equation}
A useful (and standard) identity that has been used in deriving this result is
\begin{equation}
d\bar\theta\Gamma^M d\theta d\bar\theta \Gamma_{MN}  + 
d\bar\theta\Gamma_{MN} d\theta
d\bar\theta \Gamma^M  = 0. \label{Gammaid1}
\end{equation}
The overall normalization of $b_3$ and $\delta_\epsilon B$ could be scaled
arbitrarily (including zero) as far as the present reasoning is concerned.  The
specific choice that has been made is the one that will be required later.  We
also note, for future reference, that
\begin{equation}
d {\cal H} = - db_3 = -{1\over 2}d \bar\theta \Gamma_{MN} d\theta \Pi^M \Pi^N =
-{1\over 2} d\bar\theta \psi_5^2 d\theta.
\end{equation}
where we have introduced the matrix valued one-form
\begin{equation}
\psi_5 = \Gamma_M \Pi^M_{\mu} d\sigma^{\mu}.
\end{equation}

With these choices for $G_{\hat\mu \hat\nu}$ and ${\cal H}$, we can now write
down extensions of $L_1$ and $L_3$ that have manifest 11d super-Poincar\'e
symmetry:
\begin{eqnarray}
L_1 &=& - \sqrt{-G} \sqrt{1 + z_1 + {1\over 2} z_1^2 - z_2}\nonumber
\\
L_3 &=& {1\over 8} \epsilon_{\mu\nu\rho\lambda\sigma} {G^{5\rho}\over G^{55}}
\tilde{\cal H}^{\mu\nu} \tilde{\cal H}^{\lambda\sigma},
\end{eqnarray}
where $z_1$ and $z_2$ are now formed from ${\cal H}$ instead of $H$.

The next step is to construct a supersymmetric extension of $L_2$.  This term
is the Wess--Zumino term, which can be represented as the integral of a closed
7-form $I_7$ over a region that has the 6d world volume $M_6$ as its boundary.  In
other words, 
\begin{equation}
S_2 =\int_{M_{7}} I_7 = \int_{M_{6}} \Omega_6,
\end{equation}
where $I_7 = d \Omega_6$ and  $M_6 = \partial M_7$.  The appropriate
expression for $I_7$ that reproduces $L_2$ of the purely bosonic theory is
\begin{equation}
I_7^{(B)} = -{1 \over 2} H dH = {1\over 2} H \partial_5 H d\sigma^5. \label{I7}
\end{equation}
To understand this properly, there is a point that needs to be stressed.
Namely, in adding a formal 7th dimension, the extra dimension 
is required to enter
symmetrically with the first five.  There continues to be one preferred
direction, $\sigma^5$, that is treated specially.  Correspondingly, in writing
$M_6 = \partial M_7$, the
boundary operator should not act on the $\sigma^5$ direction.  In other words,
$M_7$ should have no $\sigma^5 =$ constant faces.  It should also be noted that
this $M$ theory five-brane theory action has a
Wess--Zumino term that survives even for the bosonic truncation in a
flat space-time background. However, as we will see in the next subsection,
this feature is particular to the non-covariant formulation and is not shared
by the PST formulation in which the pieces of the action are arranged somewhat
differently.

To complete the construction of $L_2$ we must now supersymmetrize $I_7^{(B)}$.
The term $ {1\over 2} {\cal H} \partial_5 {\cal H} d\sigma^5$ achieves this,
of course, but it is no longer closed.  Additional terms should be
added such that $d I_7 = 0$, up to a total derivative in the $\sigma^5$
direction.  The result that we find is
\begin{equation}
I_7 = {1\over 2} {\cal H} \partial_5 {\cal H} d \sigma^5 - {1\over 2} {\cal H}
d\bar\theta \psi^2 d\theta - {1\over 120} d\bar\theta \psi^5 d\theta, \label{I7form}
\end{equation}
where 
\begin{equation}
\psi = \Gamma_M \Pi^M_{\hat\mu} d\sigma^{\hat\mu}
= \psi_5 + \Gamma_M \Pi^M_{5} d\sigma^{5}.
\end{equation}
When interpreting the 4-form $d\theta \psi^2 d\theta$ and 
the 7-form $d\theta\psi^5 d\theta$
it must be understood that one of the derivatives is 
required to be in the $\sigma^5$
direction.  The proof that
$dI_7$ is a total $\sigma^5$ derivative is reasonably straightforward using the identity (\ref{Gammaid1}) as well as
\begin{equation}
{1\over 6} \big(d\bar\theta \Gamma_{MNPQR} d\theta d\bar \theta \Gamma^R + d
\bar\theta \Gamma^R d\theta d\bar\theta \Gamma_{MNPQR}\big) = d \bar\theta
\Gamma_{[MN} d\theta d\bar\theta \Gamma_{PQ]}. \label{Gammaid2}
\end{equation}

Since $I_7$ is manifestly supersymmetric, it is guaranteed that $\Omega_6$ is
invariant up to a total derivative under a supersymmetry transformation.  For
most purposes an explicit formula for $L_2$ is not required.  Here we will
simply report that
\begin{equation}
L_2 = - {1\over 4} \tilde{H}^{\mu\nu} (\partial_5 B_{\mu\nu} -2
b_{\mu\nu}) + ~{\rm terms ~indep. ~of}~ B, \label{partofL2}
\end{equation}
where $b_2 = {1\over 2} b_{\mu\nu} d\sigma^\mu d\sigma^\nu$ is 
given by\footnote{This expression is equal to $b_{\mu\nu 5}$,
where $b_{\hat\mu \hat\nu \hat\rho}$ is the covariant extension of the expression given in
eq.~(\ref{b3form}).}
\begin{eqnarray}
b_2 &=& - {1 \over 2}\bar\theta \Gamma_{MN} \partial_5 \theta (dX^M dX^N + dX^M \bar\theta
\Gamma^N d\theta + {1\over 3} d\bar\theta \Gamma^M d\theta d\bar\theta \Gamma^N
d\theta)\nonumber \\
& & +{1 \over 2}\bar\theta \Gamma_{MN} d\theta (2dX^M \partial_5 X^N - \partial_5 X^M
\bar\theta \Gamma^N d\theta - dX^M \bar\theta \Gamma^N \partial_5 \theta -
{2\over 3} \bar\theta \Gamma^M d\theta \bar\theta \Gamma^N \partial_5 \theta).
\end{eqnarray}
Knowing this much of $L_2$ is sufficient to obtain the $B_{\mu\nu}$ equation of
motion.

\medskip
\section{General Coordinate Invariance}

We should now check whether the
general coordinate invariance of the bosonic theory in sect. 2.1
continues to hold after adding terms
depending on $\theta$ in the way that we have described.  As in the bosonic
case, general coordinate invariance in five directions is manifest, so only the
transformation in the $\sigma^5$ direction needs to be checked.  The
coordinates $X^M$ and $\theta$ transform as scalars, {\it i.e.},
\begin{equation}
\delta_\xi X^M = \xi \partial_5 X^M \quad {\rm and} \quad  \delta_\xi \theta = \xi
\partial_5 \theta,
\end{equation}
which implies that $G_{\hat\mu \hat\nu}$ transforms as in eq.~(\ref{Gvar}).  To specify
the proper transformation law for $B_{\mu\nu}$, we should first examine its
equation of motion. Using eq.~(\ref{partofL2}), this is
\begin{equation}
\epsilon^{\mu\nu\rho\lambda\sigma} \partial_\rho \left(K_{\lambda\sigma} -
\partial_5 B_{\lambda\sigma} + b_{\lambda\sigma}\right) = 0.
\end{equation}
The formula for $K_{\mu\nu}$ is as given in 
eqs.~(\ref{Kform1})  and (\ref{Kform2}), except that now
$L_1$ and $L_3$ of the supersymmetrized theory should be used.  This simply
amounts to replacing $H$ by ${\cal H}$ and using the supersymmetric expression
for $G_{\hat\mu \hat\nu}$.  
By the reasoning explained in ref.~\cite{jhs}, the $B$
equation of motion suggests that the appropriate
transformation formula, generalizing eq.~(\ref{Bvar}),  is
\begin{equation}
\delta_\xi B_{\mu\nu} = \xi \big(K_{\mu\nu} + b_{\mu\nu}\big).
\end{equation}
To determine $\delta_\xi {\cal H}$, one first computes that
\begin{equation}
\delta_\xi b_3 = \xi \partial_5 b_3 + b_2 d\xi.
\end{equation}
It follows that
\begin{equation}
\delta_\xi {\cal H} = d(\delta_\xi B) - \xi
\partial_5 b_3 - b_2 d\xi  = d(\xi K) - \xi Z_3, \label{Hxivar}
\end{equation}
where
\begin{equation}
Z_3 = \partial_5 b_3 - db_2. \label{Z3form}
\end{equation}
This can be made manifestly supersymmetric by noting that
\begin{equation}
Z_3 d\sigma^5 = \big(\partial_5 b_3 - db_2\big) d\sigma^5 =
-{1\over 2} d \bar\theta \psi^2 d\theta.
\end{equation}
The 4-form on the right-hand side of this equation is required to
contain one $\sigma^5$ derivative.

The important point is that the $Z_3$ term in $\delta_{\xi} {\cal H}$ has no
counterpart in the bosonic theory, so general coordinate invariance  of the
supersymmetric theory is not an immediate consequence of the corresponding
symmetry of the bosonic theory.  Let us  examine next the part of $\delta_\xi (L_1 +
L_3)$ that arises from varying ${\cal H}$, but not $G$.  It is
\begin{equation}
\delta_\xi \tilde{\cal H}^{\mu\nu} {\partial (L_1 + L_3)\over \partial \tilde{\cal
H}^{\mu\nu}} = {1\over 2} \delta_\xi \tilde{\cal H}^{\mu\nu} K_{\mu\nu}.
\end{equation}
This is conveniently characterized by the 5-form
\begin{equation}
 (d(\xi K) - \xi Z_3) K \sim -  \xi K (dK + Z_3),
\label{conv5}
\end{equation}
where $\sim$ means that a total derivative has been dropped.

Consider now the $\xi$ transformation of $L_2$.  A portion of $L_2$ was given
in eq. (\ref{partofL2}).  Representing this as a 5-form and using
\begin{equation}
\delta_\xi b_2 = \partial_5 (\xi b_2),
\end{equation}
one obtains
\begin{eqnarray}
\delta_\xi L_2 &=& -  \big(\partial_5 B - b_2\big) d
\big(\xi \big(K +b_2\big)\big) + H \partial_5 (\xi
b_2) + \ldots\nonumber \\
& \sim & \xi K (\partial_5 {\cal H} + Z_3) + {1\over 2} b_2^2 d\xi +
\ldots
\end{eqnarray}
where the dots are the contribution from varying the $H$ independent terms in $
L_2$.  The $\ldots$ terms precisely cancel the $b_2^2$ term, leaving
\begin{equation}
\delta_\xi L_2 \sim  \xi K (\partial_5 {\cal H} + Z_3). \label{L2contrib}
\end{equation}
The demonstration that the $\ldots$ terms contribute $-{1\over 2} b_2^2 d\xi $
can be made as follows. The first two terms in eq.~(\ref{I7form}) contribute
the non-$H$ pieces
\begin{equation}
{1 \over 2} b_3 \partial_5 b_3 d \sigma^5 + {1 \over 2} b_3 d \bar\theta \psi^2 d \theta,
\end{equation}
which has a non-trivial $\xi$ transformation, because of the asymmetric way in which
the $\sigma^5$ direction appears. The variation is easy to compute, and can be
expressed as the exterior derivative of $-{1\over 2} b_2^2 d\xi $, which implies that
this contributes the required variation of $L_2$.

Combining eq.~(\ref{L2contrib}) with eq.~(\ref{conv5}) leaves
\begin{equation}
\delta_{\cal H} (L_1 + L_3) + \delta_{\xi} L_2 \sim  \xi K (\partial_5
{\cal H} - dK).
\end{equation}
This must now be combined with the terms arising from varying $G_{\hat\mu
\hat\nu}$ in $L_1$ and $L_3$.  However, at this point all terms whose structure
is peculiar to the supersymmetric theory have cancelled.  The rest of the
calculation is  identical to that for the bosonic theory given in ref.~\cite{jhs} and, therefore,
need not be repeated here.

\medskip

\section{Proof of Kappa Symmetry}

\subsection{Formulation Without Manifest Covariance}

As with all other super $p$-branes of maximally supersymmetric theories, 
the world-volume theory should have 8 bosonic
and 8 fermionic physical degrees of freedom.  This requires, in particular, the existence
of a local fermionic symmetry (called kappa) that eliminates half of the
components of $\theta$.  Despite the lack of manifest general coordinate
invariance, the analysis of kappa symmetry for the $M$ theory five-brane is
very similar to that of other super $p$-branes.  As usual, we require that
\begin{equation}
\delta \bar\theta = \bar\kappa (1 - \gamma),
\end{equation}
where $\kappa(\sigma)$ is an arbitrary Majorana spinor and $\gamma$ is a
quantity (to be determined) whose square is the unit matrix.  This implies that
${1\over 2} (1 - \gamma)$ is a projection operator, and half of the components of
$\theta$ can be gauged away.  In addition, just as for all other super $p$-branes, we
require that
\begin{equation}
\delta X^M  = - \delta \bar\theta \Gamma^M \theta,
\end{equation}
so that
\begin{equation}
\delta \Pi_{\hat\mu}^M = - 2 \delta \bar\theta\Gamma^M \partial_{\hat\mu}
\theta.
\end{equation}
As in our other work \cite{aganagic}, we introduce the induced $\gamma$ matrix
\begin{equation}
\gamma_{\hat\mu} = \Pi_{\hat\mu}^M \Gamma_M,
\end{equation}
which satisfies
\begin{equation}
\{\gamma_{\hat\mu}, \gamma_{\hat\nu}\} = 2 G_{\hat\mu\hat\nu}.
\end{equation}
In this notation, the kappa variation of the metric is
\begin{equation}
\delta G_{\hat\mu\hat\nu} = - 2\delta \bar\theta (\gamma_{\hat\mu}
\partial_{\hat\nu} + \gamma_{\hat\nu} \partial_{\hat\mu})\theta.
\end{equation}

Before we can examine the symmetry of our theory, we must also specify the
kappa variation of $B_{\mu\nu}$.  This works in a way that is analogous to that
of the world-volume gauge field for D-branes.  Specifically, for the choice
\[
\delta B =  {1\over 2} \delta \bar\theta \Gamma_{MN} \theta (dX^M dX^N +
\bar\theta \Gamma^M d\theta d X^N + {1\over 3} \bar\theta \Gamma^M d \theta
\bar\theta \Gamma^N d\theta)\]
\begin{equation}
+ {1\over 2} \delta \bar\theta \Gamma^M \theta\bar\theta \Gamma_{MN} d\theta (d
X^N + {1\over 3} \bar\theta \Gamma^N d\theta),
\end{equation}
we find that most of the terms in $\delta {\cal H}$ cancel leaving
\begin{equation}
\delta {\cal H}_{\mu\nu\rho} = 6  \delta\bar\theta \gamma_{[\mu\nu}
\partial_{\rho]} \theta
\end{equation}
or, equivalently,
\begin{equation}
\delta \tilde{\cal H}^{\mu\nu} =  \epsilon^{\mu\nu\rho\lambda\sigma} \delta
\bar\theta \gamma_{\rho\lambda} \partial_\sigma \theta.
\end{equation}
Since we now have the complete theory and all the field transformations, it is
just a matter of computation to check the symmetry.

Before plunging into the details of the calculation, it is helpful to sketch
the general strategy that will be employed.  It turns out to be convenient to
consider $L_2$ and $L_3$ together and to write their kappa variation in the
form
\begin{equation}
\delta (L_2 + L_3) = {1\over 2} \delta \bar\theta T^{\hat\mu}
\partial_{\hat\mu} \theta.
\end{equation}
The variation of $L_1$ is represented in a similar manner:
\begin{equation}
\delta L_1 = - {1\over 2L_1} \delta\bar\theta U^{\hat\mu} \partial_{\hat\mu}
\theta.
\end{equation}
Then, in order that $\delta \bar\theta = \bar\kappa (1 - \gamma)$ should be a
symmetry, we require that altogether
\begin{equation}
\delta (L_1  + L_2 + L_3) = {1\over 2} \delta \bar\theta (1 + \gamma)
T^{\hat\mu} \partial_{\hat\mu} \theta,
\end{equation}
which is achieved if
\begin{equation}
U^{\hat\mu} = \rho T^{\hat\mu},
\end{equation}
where
\begin{equation}
\rho = - \gamma L_1 = \gamma\sqrt{-G} \sqrt{1 + z_1 + {1\over 2} z_1^2 - z_2}.
\end{equation}
This implies that
\begin{equation}
\rho^2 = - G \big( 1 + z_1 + {1\over 2} z_1^2 - z_2\big). \label{rho2}
\end{equation}
We must  vary the Lagrangian to find $T^{\hat\mu}$ and $U^{\hat\mu}$, and then determine
$\rho$ with the proper square and show that $U^{\hat\mu} = \rho
T^{\hat\mu}$.  This is all straightforward, but it needs to be done carefully.

Since the $\sigma^5$ direction appears asymmetrically in the Lagrangian, the
analysis of $U^{\hat\mu} = \rho T^{\hat\mu}$ is naturally split into two
separate problems, corresponding to $\hat\mu = 5$ and $\hat\mu \not= 5$.  The
$\hat\mu = 5$ case is the easier of the two, so let us begin with that.  We
must examine where we can get $\partial_5 \theta$'s.  The variations of
$B_{\mu\nu}$ and $G_{\mu\nu}$ do not give any.  Therefore, in varying $L_1$,
the variations of $z_1$ and $z_2$ do not contribute.  The only contribution
comes from
\begin{equation}
\delta \sqrt{-G} = - 2 \sqrt{-G} \delta \bar\theta \gamma^{\hat\mu}
\partial_{\hat\mu} \theta,
\end{equation}
where, of course, $\gamma^{\hat\mu} = G^{\hat\mu\hat\nu} \gamma_{\hat\nu}$.
Thus
\begin{equation}
U^5 = - 4 \rho^2 \gamma^5.
\end{equation}
To determine $T^5$ we must vary $L_2 + L_3$.  Using the identity
\begin{equation}
\delta\left({G^{5\rho}\over G^{55}}\right) = 2 {G_5^{\eta\rho}
G^{5\hat\mu}\over G^{55}} \delta \bar\theta (\gamma_{\hat\mu} \partial_\eta +
\gamma_\eta \partial_{\hat\mu})\theta,
\end{equation}
the relevant piece of $\delta L_3$ is
\begin{equation}
{1\over 4} \epsilon_{\mu\nu\rho\lambda\sigma} G_5^{\eta\rho} \delta \bar\theta
\gamma_\eta \partial_5 \theta \tilde{\cal H}^{\mu\nu} \tilde{\cal
H}^{\lambda\sigma},
\end{equation}
which contributes 
\begin{equation}
T_2^5 = {1\over 2} \epsilon_{\mu\nu\rho\lambda\sigma}
G_5^{\eta\rho} \gamma_\eta \tilde{\cal H}^{\mu\nu} \tilde{\cal
H}^{\lambda\sigma} \label{T25}
\end{equation}
 to $T^5$. (The subscript on $T$ represents the power of ${\cal H}$.)

The variation of the Wess--Zumino term $S_2$ is
\begin{equation}
\delta S_2 =  \int ({\cal H} \delta \bar\theta \psi^2 d\theta - {1\over 60}
\delta \bar\theta \psi^5 d\theta),
\end{equation}
a result that is obtained by expressing $\delta I_7$ as a total differential.
This determines $T_0^5 + T_1^5$, with
\begin{equation}
T_0^5 = - {1\over 30} \epsilon^{\mu_{1} \ldots \mu_{5}} \gamma_{\mu_{1} \ldots
\mu_{5}} = - 4 \bar\gamma \gamma^5, \label{T05}
\end{equation}
where we have introduced
\begin{equation}
\bar\gamma = \gamma_{012345},
\end{equation}
which satisfies $(\bar\gamma)^2 = - G$.  The ${\cal H}$ linear term is
\begin{equation}
T_1^5 = - 2 \tilde{\cal H}^{\mu\nu} \gamma_{\mu\nu}. \label{T15}
\end{equation}
Combining these results with
\begin{equation}
U^5 = - 4 \rho^2 \gamma^5 = \rho T^5,
\end{equation}
we infer that $T^5 = - 4 \rho \gamma^5$, where
\begin{equation}
\rho = \bar\gamma + {1\over 2G^{55}} \tilde{\cal H}^{\nu\rho} \gamma_{\nu\rho}
\gamma^5 - {1\over 8G^{55}} \epsilon_{\mu\nu\rho\lambda\sigma} \tilde{\cal
H}^{\mu\nu} \tilde{\cal H}^{\rho\lambda} \gamma^{\sigma 5}. \label{rhoform}
\end{equation}
To obtain the ${\cal H}^2$ term we have used the identity
\begin{equation}
G_5^{\eta\sigma} \gamma_\eta = \gamma^\sigma - {G^{\sigma 5}\over G^{55}}
\gamma^5, \label{G5id1}
\end{equation}
from which it follows that
\begin{equation}
G_5^{\eta\sigma} \gamma_\eta \gamma^5 = \gamma^{\sigma 5}. \label{G5id2}
\end{equation}
If our reasoning is correct, this expression for $\rho$ should have the square
given in eq. (\ref{rho2}).  This fact is verified in Appendix A.

To complete the proof of kappa symmetry, we must find $U^\mu$ and $T^\mu$
and show that $U^\mu = \rho T^\mu$.  Separating powers of ${\cal H}$, as
above, the variation of $L_2$ contributes to $T_0^\mu$ and $T_1^\mu$ while the
variation of $L_3$ contributes to $T_1^\mu$ and $T_2^\mu$.  Altogether, we find
that
\begin{eqnarray}
T_0^\mu &=& - 4 \bar\gamma \gamma^\mu\nonumber \\
T_1^\mu &=& -{2\over G^{55}} (G^{5\mu} \tilde{\cal H}^{\nu\rho} \gamma_{\nu\rho}
+ 2 \tilde{\cal H}^{\mu\nu} \gamma_\nu \gamma^5)\nonumber \\
T_2^\mu &=& {1\over 2G^{55}} \epsilon_{\eta\nu\rho\lambda\sigma} \tilde{\cal
H}^{\nu\rho} \tilde{\cal H}^{\lambda\sigma} (G^{5\mu} G_5^{\eta\zeta}
\gamma_\zeta + G_5^{\mu\eta} \gamma^5). \label{Tforms}
\end{eqnarray}
The variation of $L_1$ determines $U^\mu = \sum_{n = 0}^4 U_n^\mu$, where
\begin{eqnarray}
U_0^\mu &=& 4 G\gamma^\mu\nonumber \\
U_1^\mu &=& -{1\over G^{55}} \epsilon^{\mu\nu\rho\lambda\sigma}
\gamma_{\lambda\sigma} (G\tilde{\cal H} G)_{\nu\rho}\nonumber \\
U_2^\mu &=& - {4\over G^{55}} \gamma_\nu (\tilde{\cal H} G \tilde{\cal
H})^{\mu\nu} - {2\over (G^{55})^2} G^{5\mu} \gamma^5 {\rm tr} (G\tilde{\cal H}
G\tilde{\cal H})\nonumber \\
U_3^\mu &=&  {1\over G(G^{55})^2} \epsilon^{\mu\nu\rho\lambda\sigma}
\gamma_{\lambda\sigma} \left({1\over 2} (G\tilde{\cal H} G)_{\nu\rho} {\rm tr}
(G\tilde{\cal H} G\tilde{\cal H}) - (G\tilde{\cal H} G\tilde{\cal H}
G\tilde{\cal H} G)_{\nu\rho}\right)\nonumber \\
U_4^\mu &=& {4\over G(G^{55})^2} \gamma_\nu \left({1\over 2}(\tilde{\cal H}
G \tilde{\cal H})^{\mu\nu} {\rm tr} (G {\cal H} G {\cal H}) - (\tilde{\cal H} G
\tilde{\cal H} G \tilde{\cal H} G \tilde{\cal H})^{\mu\nu}\right)\nonumber \\
&+& {2\over G(G^{55})^2} \left(G^{\mu 5} \gamma^5 - {1\over 2} G^{55}
\gamma^\mu \right) \left({1\over 2} ({\rm tr} (G \tilde{\cal H} G \tilde{\cal H}))^2
- {\rm tr} ( G{\cal H}G{\cal H}G{\cal H}G{\cal H})\right). \label{Uforms}
\end{eqnarray}
The demonstration that $U^\mu = \rho T^\mu$ is presented in Appendix B.

In conclusion, we have shown that the theory specified by $L_1 + L_2 + L_3$ has
all the desired symmetries: global 11d super--Poincar\'e symmetry, 
general coordinate invariance, and local  kappa symmetry.

\subsection{Supersymmetric Theory in the PST Formulation}

The supersymmetric theory that we have just presented can be recast in a
manifestly general covariant form, using the PST formalism, just as we did for
the bosonic theory in sect. 2.2.  In order to keep the notation from being
too cumbersome, in this section (and only in this section) indices $\mu$, $\nu$,
etc., take six values, ({\it i.e.}, we drop the hats used until now).  Also the label
``cov.'' is dropped.  Thus, upon supersymmetrization, eq.~(\ref{Mcov}), for example,
becomes
\begin{equation}
M_{\mu\nu} = G_{\mu\nu} + i {G_{\mu\rho} G_{\nu\lambda}\over\sqrt{-G (\partial
a)^2}} \tilde{\cal H}^{\rho\lambda},
\end{equation}
where
\begin{equation}
\tilde{\cal H}^{\rho\lambda} = {1\over 6}
\epsilon^{\rho\lambda\mu\nu\sigma\tau} {\cal H}_{\mu\nu\sigma} \partial_\tau a.
\end{equation}
Also, $G_{\mu\nu}$ is constructed as in eqs.~(\ref{Pidef}) and 
(\ref{susymetric}), and ${\cal H} = H - b_3$ is extended to six dimensions.
In this notation the supersymmetric theory is given by $L=L_1 + L' + L_{WZ}$,
where
\begin{eqnarray}
L_1 &=& - \sqrt{- {\rm det}\, M_{\mu\nu}}\nonumber \\
L' &=& - {1\over 4(\partial a)^2} \tilde{\cal H}^{\mu\nu} {\cal H}_{\mu\nu\rho}
G^{\rho\lambda} \partial_\lambda a\\
S_{WZ} &=& \int \Omega_6.\nonumber 
\end{eqnarray}
$L_1$ can again be recast in the form
\begin{equation}
L_1 = - \sqrt{-G} \sqrt{1 + z_1 + {1\over 2} z_1^2 - z_2},
\end{equation}
where now $z_1$ and $z_2$ are the obvious covariant counterparts of 
those in eq.~(\ref{zdefs}).  The
Wess--Zumino term is again characterized by a seven-form $I_7 = d \Omega_6$,
where now
\begin{equation}
I_7 = - {1\over 4} {\cal H} d \bar\theta \psi^2 d\theta - {1\over 120} d
\bar\theta \psi^5 d\theta. \label{I7cov}
\end{equation}
It is easy to check that $dI_7 = 0$ using 
eqs.~(\ref{Gammaid1}) and (\ref{Gammaid2}).  Global $\epsilon$
supersymmetry and local reparametrization symmetry are manifest in these
formulas.  Note that neither the metric $G_{\mu\nu}$ nor the scalar field $a$
occur in $L_{WZ}$.

When one chooses the gauge $a = \sigma^5$ and $B_{\mu 5} = 0$, the Lagrangian
given above reduces to the one in sect. 3.  The way this happens is somewhat
non-trivial.  The point is that $L'$ reduces to $L_3$ and a portion of the
non-covariant Wess--Zumino term $L_2$.  Specifically, in the
gauge-fixed theory the sum over the index $\rho$ in the formula for $L'$ can
be separated into $\rho = 5$ and $\rho \not= 5$ terms.  The $\rho \not= 5$ term
accounts for $L_3$ of the gauge-fixed theory, while the $\rho = 5$ term
accounts for the ${\cal H}^2$ piece of $L_2$ and a portion of the ${\cal H}$
piece.  In particular, this accounts for why the coefficient 
of the ${\cal H}$ linear term in
eq.~(\ref{I7cov}) differs from that in eq.~(\ref{I7form}).

The proof of kappa symmetry in the PST formulation
works as before (with $\delta a = 0$), so we will
not repeat the argument.\footnote{Also, D. Sorokin informs us
that it will appear soon in a paper by him and collaborators.} 
The covariant extension of eq.~(\ref{rhoform}) is
\begin{equation}
\rho = \bar\gamma + {1\over 2(\partial a)^2} \tilde{\cal H}^{\nu\rho}
\gamma_{\nu\rho} \gamma^\lambda \partial_\lambda a - {1\over 16 (\partial a)^2}
\epsilon_{\mu\nu\rho\lambda\sigma\tau} \tilde{\cal H}^{\mu\nu} \tilde{\cal
H}^{\rho\lambda} \gamma^{\sigma\tau}.
\end{equation}
The demonstration that $\rho^2 = - {\rm det}\, M_{\mu\nu}$ is essentially the same as
in Appendix A.  The covariant formula for $T^\mu = T_0^\mu + T_1^\mu + T_2^\mu$ is
given by
\begin{eqnarray}
T_0^\mu &=& - 4 \bar\gamma \gamma^\mu\nonumber \\
T_1^\mu &=& -{2\over (\partial a)^2} \tilde{\cal H}^{\nu\rho} (\gamma_{\nu\rho}
G^{\mu\lambda} - 2 \delta_\rho^\mu \gamma_\nu \gamma^\lambda)\partial_\lambda
a\nonumber \\
T_2^\mu &=&  - {1\over (\partial a)^2} \tilde{\cal H}^{\eta\nu} {\cal
H}_{\eta\nu\rho} (\gamma^\rho G^{\lambda\mu} + \gamma^\lambda G^{\rho\mu})
\partial_\lambda a \nonumber \\
&& +{2\over [(\partial a)^2]^2} \tilde{\cal H}^{\eta\nu} {\cal H}_{\eta\nu\rho}
G^{\rho\lambda} \partial_{\lambda} a \gamma^\sigma \partial_\sigma a
G^{\mu\zeta} \partial_\zeta a.
\end{eqnarray}
In the $B_{\mu 5} = 0,$ $ a = \sigma^5$ gauge, these expressions reduce to the
formulas $T^5$ and $T^\mu$ given in 
eqs.~(\ref{T25}), (\ref{T05}), (\ref{T15}), and (\ref{Tforms}).  
The proof of kappa symmetry works
essentially the same as before.  

\medskip

\section{Double Dimensional Reduction}

As is now well-known, when one of the ten spatial dimensions
of M theory is a small circle of radius $R$, the theory can be
reinterpreted as Type IIA string theory in ten dimensions with
string coupling constant proportional to $R^{3/2}$~\cite{townsend3,witten3}. 
The five-brane
of M theory can then give rise to either a five-brane  or a four-brane
of Type IIA string theory depending on whether or not it wraps around
the circular dimension. Here we wish to focus on the case that it
does wrap (once) so that one obtains a four-brane. This case is called
``double dimensional reduction,'' because the dimension of the brane
and the dimension of the ambient space-time have been reduced by one
at the same time. (The first example of this type to be studied was the
double dimensional reduction of the M theory two-brane, which gives the
Type IIA fundamental string~\cite{duff2}.) The known 4-brane
of Type IIA string theory is, in fact, a D-brane, which implies 
that its world-volume theory contains an abelian vector gauge field.
However, the five-brane theory that we have constructed contains an
antisymmetric tensor gauge field, which remains one even after the
reduction. However, as we will show elsewhere~\cite{aganagic2}, the D4-brane action
and the 4-brane with antisymmetric tensor gauge field obtained below,
are related by a world-volume duality transformation. This is analogous
to the relationship between the M2-brane 
and the D2-brane~\cite{duff,townsend,schmidhuber}.

The covariant action for the dual D4-brane 
in ten dimensions can be obtained from the 
M theory five-brane action by setting
\begin{equation}\label{red.-eq}
X^{11} = \sigma^5
\end{equation}
and then dropping all dependence on $\sigma^5$, {\it i.e.}, extracting the zeroth
Fourier mode.   Doing this gives
\begin{eqnarray}
\psi &\rightarrow & \psi + C \Gamma_{11}  \\
G_{\mu\nu} &\rightarrow & G_{\mu\nu} + C_\mu C_\nu \\
b_3 &\rightarrow & C_3 ,
\end{eqnarray}
where
\begin{equation}
C _{\mu} =  - \bar\theta \Gamma^{11} \partial_{\mu}\theta
\end{equation} 
is the part of $\Pi^{11}_{\mu}$ that survives. $C$ and
\begin{equation}
C_3 =  b_3 + {1 \over 2} \bar\theta \Gamma_{11} \Gamma_n d \theta
\bar\theta \Gamma^{11} d \theta 
( d X^n + {2 \over 3} \bar\theta \Gamma^n d \theta)
\end{equation} 
enter in the D4-brane Wess-Zumino term.
In these formulae quantities on the left (right) of the arrow have
target space indices summed on 11 (10) values (e.g., $\psi = \Gamma_M
\Pi^M$ on the L.H.S., $\psi = \Gamma_m \Pi^m$ on the R.H.S.,
where $m = 0,1,\ldots,9$ and $M=(m,11)$).  Also,
\begin{eqnarray}
G = \det G_{\hat\mu\hat\nu} & \rightarrow &  G = \det G_{\mu\nu}\nonumber\\
G_5 = \det G_{\mu\nu} &\rightarrow & \det (G_{\mu\nu} + C_\mu C_\nu) = G (1 +
C^2),
\end{eqnarray}
where
\begin{equation}
C^2 \equiv G^{\mu\nu} C_\mu C_\nu.
\end{equation}

One can analyze the double dimensional reduction of the action.
A straightforward calculation shows that
\begin{equation}
\det \left(G_{\hat\mu\hat\nu} + i {G_{\hat\mu \rho} G_{\hat\nu \lambda}
\tilde{\cal H}^{\rho\lambda}\over\sqrt{-G_5}}\right) \rightarrow \det
\left(G_{\mu\nu} + i {G_{\mu\rho} G_{\nu\lambda} \tilde{\cal
H}^{\rho\lambda}\over\sqrt{-G (1 + C^2)}} + Y_\mu Y_\nu \right)
\end{equation}
with
\begin{equation}
Y_\mu \equiv i {G_{\mu\rho} \tilde{\cal H}^{\rho\lambda} 
C_\lambda \over\sqrt{-G (1 + C^2)}},
\end{equation}
which gives the double-dimensionally reduced version of ${  L}_1$.
For ${  L}_3$ the answer is:
\begin{equation}
{  L}_3 = {1\over 8} \epsilon_{\mu\nu\rho\lambda\sigma} {G^{5\rho}\over
G^{55}} \tilde{\cal H}^{\mu\nu} \tilde{\cal H}^{\lambda\sigma} \rightarrow -
{1\over 8} \epsilon_{\mu\nu\rho\lambda\sigma} {C^\rho\over 1 + C^2}
\tilde{\cal H}^{\mu\nu} \tilde{\cal H}^{\lambda\sigma}.
\end{equation}
The Wess--Zumino term is given by the reduction 
\begin{equation}
I_7 \rightarrow  I_6 = - {1\over 4!} d\bar\theta
\Gamma_{11} \psi^4 d \theta + {\cal H} d
\bar\theta \Gamma_{11} \psi d \theta.
\end{equation}

Under double dimensional reduction
\begin{equation}
d {\cal H} = - {1\over 2} d \bar\theta \psi^2 d\theta \rightarrow - {1\over 2}
d \bar\theta \psi^2 d \theta + d \bar\theta \Gamma_{11} \psi d \theta C,
\end{equation}
whose supersymmetry variation is
\begin{equation}
\delta_\epsilon d {\cal H} \rightarrow d \bar \theta \Gamma_{11} \psi d
\theta \bar\epsilon \Gamma^{11} d \theta.
\end{equation}
From this one can infer that
\begin{equation}\label{susy}
\delta_\epsilon {\cal H} \rightarrow  (\bar\epsilon \Gamma^{11} \theta) d
\bar\theta \Gamma_{11} \psi d \theta + {\rm total\,\, derivative}.
\end{equation}
It is an interesting fact that, after the double dimensional reduction, ${\cal H}$
is no longer invariant under supersymmetry. We will show below that
the formula has a simple interpretation, which ensures that the reduced theory
is supersymmetric.
The kappa variations of the doubly dimensionally reduced theory can be
analyzed in a similar manner. One finds that
\begin{equation}\label{kappa}
\delta {\cal H} = - \delta \bar\theta \psi^2 d \theta \rightarrow - \delta
\bar\theta \psi^2 d \theta + 2 \delta \bar\theta \Gamma_{11} \psi d \theta C.
\end{equation}

In order to preserve the gauge choice~(\ref{red.-eq}), 
both the supersymmetry and the $\kappa$
variations  of the 4-brane fields must include
compensating $\sigma^5$ general coordinate transformations:
\begin{eqnarray}
0 &=& \delta_\epsilon X^{11} + \xi_\epsilon^{\hat\mu} \partial_{\hat\mu}
X^{11} = \bar\epsilon \Gamma^{11} \theta + \xi_\epsilon \nonumber \\
&\Rightarrow & \xi_\epsilon = - \bar\epsilon \Gamma^{11} \theta \nonumber
\\
0 &=& \delta X^{11} + \xi_{\kappa}^{\hat\mu} \partial_{\hat\mu} X^{11} = - \delta
\bar\theta \Gamma^{11} \theta + \xi_{\kappa}\nonumber \\
&\Rightarrow& \xi_\kappa = \delta \bar\theta \Gamma^{11} \theta.
\end{eqnarray}
Upon double dimensional reduction the induced general coordinate
transformation parameter $\xi$ only appears in the
quantities (see eqs.~(\ref{Hxivar}) and (\ref{Z3form}))
\begin{equation}\label{gct}
\delta_\xi {\cal H} = d (\xi K) + \xi db_2
\end{equation}
and
\begin{equation}
\delta_\xi C_\mu = \partial_\mu \xi.
\end{equation} 
The supersymmetry variations of $C$ and ${\cal H}$ are
entirely given by the induced $\sigma^5$ general coordinate transformation.
Therefore supersymmetry of the theory after double dimensional reduction is a
consequence of both the supersymmetry and the general coordinate invariance of
the original 6d theory.
As a consistency check, one can show that  eq.~(\ref{gct}) with $\xi = \xi_\epsilon$
reproduces eq.~(\ref{susy}).  Kappa symmetry works similarly:
\begin{equation}
\delta C_\mu = - \delta \bar\theta \Gamma^{11} \partial_\mu \theta -
\bar\theta \Gamma^{11} \partial_\mu \delta \theta = \partial_\mu \xi_\kappa - 2
\delta \theta \Gamma^{11} \partial_\mu \theta,
\end{equation}
where the second term is the remnant of the $\kappa$ variation of $G_{\mu 5}$.
Looking at $\delta (d{\cal H})$ we can compute
\begin{equation}\label{IIA}
\delta {\cal H} = - \delta \bar\theta \psi^2 d \theta + 2 \delta \bar\theta
\Gamma_{11} \psi d \theta C - (\delta \bar\theta \Gamma^{11} \theta) d \bar\theta
\Gamma_{11} \psi d \theta + {\rm total\,  derivative},
\end{equation}
which is reproduced by combining eqs.~(\ref{kappa}) and (\ref{gct}) for $\xi = \xi_\kappa$.

\medskip
\section{Discussion}

This paper has presented the world-volume action of the M theory
five-brane in a flat 11d background. The required global and local symmetries
have been verified in detail using a formulation in which one world-volume
direction is treated differently from the others. The corresponding results in
the manifestly covariant PST formulation have also been presented. Although
we have not done it, we expect that it would be reasonably straightforward to
extend the results to an arbitrary background, as has been done for D-branes
in refs.~\cite{cederwall2,bergshoeff1}.  All the considerations
in this paper have been classical, but there are undoubtedly various quantum implications.
In fact, it has been suggested recently that certain supersymmetric 6d theories
can have non-trivial renormalization group fixed points~\cite{seiberg}. 
Perhaps our five-brane action is of this type.

The five-brane world-volume theory has a solitonic solution~\cite{perry}
that describes a finite-tension self-dual string of the type discussed in~\cite{witten1}.
We think that it will be very interesting to study this string and its excitation
spectrum, which could then be compared to the spectrum conjectured in~\cite{dijkgraaf}.
It is curious that the five-brane, which itself arises as a soliton of the 11d
theory, has its own solitons. Upon double dimensional reduction to the IIA
4-brane, as discussed in sect. 6, the self-dual string can either wrap or not wrap.
This reflects the fact that the D4-brane has both point-like and string-like
solitons, which are electric-magnetic duals of one another. The point-like
solitons can also be viewed as describing bound states of D4-branes and D0-branes
with the D0-brane charge representing momentum in the compact dimension.
The string-like solitons do not appear to have an analogous interpretation.

Another direction that we think deserves to be explored is how the
M5-brane should be described in the background that describes the
$E_8\times E_8$ theory~\cite{horava}. The 5-brane in such a background
will have half as
much supersymmetry as we have described, corresponding to $N=1$ in 10d. 
More significantly,
it should have a soliton solution that describes a ``heterotic'' self-dual string.
The gauge group, whose currents would appear as left-movers, should be
$E_8$ ~\cite{ganor1,ganor2}. It would also be interesting to explore how wrapping
M5-branes on suitable 2-cycles gives rise to Seiberg--Witten theories in the
unwrapped dimensions~\cite{klemm}.

\newpage
\section*{Appendix A -- Evaluation of $\rho^2$}

This appendix will show that $\rho^2 = - G(1+z_{1}+ \frac{1}{2} z_{1}^2 -z_{2})$,
where
\begin{equation}
\rho = \bar{\gamma} +
\frac{1}{2G^{55}}\tilde{\cal H}^{\nu\rho}\gamma_{\nu\rho}\gamma^{5}-
\frac{1}{8 G^{55}} \epsilon_{\mu\nu\rho\lambda\sigma}
  \tilde{\cal H} ^{\mu\nu} \tilde{\cal H}^{\rho\lambda}\gamma^{\sigma5}.
\end{equation}
It is convenient to rewrite $\rho_{2}$ (the subscript refers to the order in $\cal H$) as
\begin{equation}
\rho_{2}=   \frac{1}{8 G_{5}}  \bar{\gamma}  
\gamma_{\mu\nu\rho\lambda}
            \tilde{\cal H} ^{\mu\nu} \tilde{\cal H}^{\rho\lambda},
\end{equation}
where we have used
\begin{equation}\label{eq:g6k}
\frac{1}{k!}\epsilon_{\hat{\mu}_{1}\ldots\hat{\mu}_{6}}
\gamma^{\hat{\mu}_{1}\ldots\hat{\mu}_{k}}=
\frac{1}{G}(-1)^{\frac{k(k+1)}{2}}
\bar{\gamma}\gamma_{\hat{\mu}_{k+1}\ldots\hat{\mu}_{6}}. \label{appaform}
\end{equation}
The matrix
$\bar{\gamma}$ anticommutes with all $\gamma^{\mu}$'s, so
 $\{ \rho_{0},\rho_{1} \} = 0$  and $[ \rho_{0},\rho_{2} ] = 0$. Furthermore, 
\begin{equation}
\{ \rho_{1},\rho_{2} \} \sim
 [\gamma_{\alpha\beta},\gamma_{\mu\nu\rho\sigma}]
\tilde{\cal H}^{\alpha\beta}\tilde{\cal H}^{\mu\nu}
\tilde{\cal H}^{\rho\sigma}=0
\end{equation}
as the commutator is antisymmetric over six 5-valued indices. 
Thus,
\begin{equation}
\rho^2 = \rho_{0}^2+\rho_{1}^2+2\rho_{0}\rho_{2}+\rho_{2}^2.
\end{equation}

We know already that $\rho_{0}^2=-G$ and
$\rho_{0}\rho_{2} = -\frac{1}{8G^{55}}
\tilde{\cal H}^{\mu\nu}\tilde{\cal H}^{\rho\sigma}
\gamma_{\mu\nu\rho\sigma}.$ So we need
\begin{eqnarray*}
\rho_{1}^2&=& \frac{1}{4G^{55}} \tilde{\cal H}^{\mu\nu}
                                   \tilde{\cal H}^{\rho\sigma}
                                   \gamma_{\mu\nu}\gamma_{\rho\sigma}
          = \frac{1}{4G^{55}} \tilde{\cal H}^{\mu\nu}
                                \tilde{\cal H}^{\rho\sigma}
               (\gamma_{\mu\nu\rho\sigma} 
             - 2G_{\mu\rho}G_{\nu\sigma})\\
         &=& \frac{1}{4G^{55}}[
\tilde{\cal H}^{\mu\nu}\tilde{\cal H}^{\rho\sigma}
\gamma_{\mu\nu\rho\sigma}
               + 2 {\rm tr}\,(\tilde{\cal H}^2)],
\end{eqnarray*}
where ${\rm tr}\,(\tilde{\cal H}^2)$ represents
 ${\rm tr}\,(G\tilde{\cal H}G\tilde{\cal H})$.
Thus, $\rho_{1}^2 +2\rho_{0}\rho_{2}=- G z_1$.
Finally,
\begin{eqnarray*}
\rho_{2}^2= -\frac{G}{64 G_{5}^2} 
\tilde{\cal H}^{\mu_1\mu_2}  \tilde{\cal H}^{\mu_3\mu_4}  
\tilde{\cal H}^{\nu_1\nu_2}  \tilde{\cal H}^{\nu_3\nu_4}
\gamma_{\mu_1\mu_2\mu_3\mu_4}\gamma_{\nu_1\nu_2\nu_3\nu_4}.
\end{eqnarray*}
In the multiplication of gamma matrices\footnote{A useful generalisation 
of the relation
 $\gamma_{\mu_1 \ldots \mu_m} \gamma_{\mu} = 
\gamma_{\mu_1 \ldots \mu_m\mu} + 
 m \gamma_{[\mu_1 \ldots \mu_{m-1}} G_{\mu_m] \mu}$ is
\[
 \gamma_{\mu_1 \ldots \mu_m} \gamma_{\nu_1 \ldots \nu_n} = 
    \sum_{k=0}^{min(m,n)} C_{k}^{mn}
 \gamma_{\mu_1 \ldots \mu_{m-k}\nu_1 \ldots \nu_{n-k}}
 G_{\mu_{m-k+1}\nu_{n-k+1}} \ldots G_{\mu_m\nu_n},
\]
where $C_{k}^{mn}\equiv(-1)^{kn+\frac{k(k+1)}{2}}
 k! \left(\stackrel{m}{k}\right) \left(\stackrel{n}{k}\right)$.
The terms in the sum are antisymmetrized over all 
$\mu$'s and $\nu$'s separately.} 
one can argue that the only terms that
contribute after contraction with the ${\cal H}$'s are effectively
\begin{eqnarray*}
\gamma_{\mu_1\mu_2\mu_3\mu_4}\gamma_{\nu_1\nu_2\nu_3\nu_4}\sim
8G_{\mu_1\nu_1}G_{\mu_2\nu_2}G_{\mu_3\nu_3}G_{\mu_4\nu_4}
-16G_{\mu_1\nu_2}G_{\mu_2\nu_3}G_{\mu_3\nu_4}G_{\mu_4\nu_1},
\end{eqnarray*}
 and thus
\begin{equation}
\rho_{2}^2= - \frac{G}{4G_5^2}\big(\frac{1}{2}{\rm tr}\,(\tilde{\cal H}^2)^2 - 
{\rm tr}\,(\tilde{\cal H}^4)\big)
 =-G\big(\frac{1}{2}z_1^2 - z_2\big).
\end{equation}
Collecting all the terms, we obtain the desired relation:
\begin{equation}
\rho^2 = -G\big(1+z_1+\frac{1}{2}z_1^2 -z_2\big).
\end{equation}

\medskip
\section*{Appendix B -- Evaluation of $\rho T^{\mu}$}

We wish to demonstrate that $\rho T^{\m}=U^{\m}$, 
where $\rho$, $T^{\mu}$, and $U^{\m}$ are given by 
eqs.~(\ref{rhoform}), (\ref{Tforms}), and (\ref{Uforms}), respectively.
The calculation is somewhat messy, so we proceed order by order in $\cal H$.

To zeroth order, $\rho_{0}=\bar{\g}$ and $T_{0}^{\m}=-4\bar{\g}\g^{\m}$
give $\rho_{0} T_{0}^{\m}=4G\g^{\m} = U_{0}^{\m}$. 
The linear order contribution comes from 
$(\rho T^{\m})_{1} = \q_{1} T_{0}^{\m}+\q_{0} T_{1}^{\m}$, where
\begin{equation}
\rho_{1}= \frac{1}{2G^{55}}
                       \hh^{\n\q}
                       {\g_{\n\q}}^{5}, \ \ \ \
           T^{\m}_{1}= - \frac{2}{G^{55}} \hh^{\n\q}
         (G^{5\m}\g_{\n\q}+2{G^{\m}}_{\n}{\g_{\q}}^{5}).
\end{equation}
Since
\begin{equation}\nonumber
\rho_{1} T_{0}^{\m} =  - \frac{2}{G^{55}}\hh^{\n\q}
                             {\g_{\n\q}}^{5} \bar{\g}\g^{\m}
                    = \frac{2}{G^{55}}\hh^{\n\q}\bar{\g}({\g_{\n\q}}^{5\m}
                         +G^{5\m}\g_{\n\q}+2{G^{\m}}_{\n}{\g_{\q}}^{5}),
\end{equation}
we obtain
\begin{equation}
(\rho T^{\m})_{1}=\frac{2}{G^{55}}\hh^{\n\q}\bar{\g}{\g_{\n\q}}^{5\m}=
     -\frac{1}{G^{55}}\epsilon^{\n\q\la\s\m}\g_{\n\q}(G \hh G)_{\la\s},
\end{equation}
where eq.~(\ref{appaform}) has been used in obtaining the second equality. Thus,
$ (\q T^{\m})_{1}=U^{\m}_{1}$.

The higher-order calculations somewhat simplify 
if one rewrites $\q_{2}$ as 
\begin{equation}
 \rho_{2}  =\frac{1}{8G_{5}}\hh^{\n\q}\hh^{\la\s}\bar{\g}\g_{\n\q\la\s}
\end{equation}
and $T_{2}^{\m}$ as
\begin{eqnarray}
T_{2}^{\m}&=&\frac{1}{2G^{55}}\hh^{\n\q}\hh^{\la\s}\epsilon_{\eta\n\q\la\s}
              (G^{\m5}G_{5}^{\eta\zeta}\g_{\zeta}+G_{5}^{\m\eta}\g^{5})\\
          &=&\frac{1}{2G_{5}}\hh^{\n\q}\hh^{\la\s}\bar{\g}
       ({\g_{\n\q\la\s}}^{\m} -2\frac{G^{\m5}}{G^{55}}{\g_{\n\q\la\s}}^{5}),
\end{eqnarray}
 using eqs.~(\ref{G5id1}) and (\ref{G5id2}).
In quadratic order,
\begin{equation}\label{eq:2}
(\rho T^{\m})_{2}=\q_{0} T^{\m}_{2}+\q_{1} T^{\m}_{1}+\q_{2} T^{\m}_{0}.
\end{equation}
If we factor out $(\hh^{\n\q}\hh^{\la\s})$ as a common factor in all terms,
\begin{eqnarray*}
&\rho_{0} T^{\m}_{2}&\!\sim - \frac{1}{2G^{55}}
   ({\g_{\n\q\la\s}}^{\m}-\frac{2G^{5\m}}{G^{55}}{\g_{\n\q\la\s}}^5 )\\
&\rho_{1}T^{\m}_{1}&\! \sim -\frac{1}{(G^{55})^2}  
     {\g_{\n\q}}^5[G^{5\m}\g_{\la\s}+ 2{G^{\m}}_{\la}{\g_{\s}}^{5}]=\\
&&      - \frac{1}{(G^{55})^{2}} [G^{5\m}({\g_{\n\q\la\s}}^{5} -
           2G_{\la\n}G_{\s\q} \g^5)
           - 2G^{55}{G^{\m} }_{\la}( \g_{\n\q\s}
           + 2\g_{\n}G_{\s\q})]\\
&\rho_{2} T^{\m}_{0}& \!\sim  - \frac{1}{2G_{5}}   
      \bar{\g}\g_{\n\q\la\s}\bar{\g}\g^{\m}=
  \frac{1}{2G^{55}}({\g_{\n\q\la\s}}^{\m}+4\g_{\n\q\la}{G_{\s}}^{\m}).
\end{eqnarray*}
Combining these contributions and reinstating $(\hh^{\n\q}\hh^{\la\s})$ gives
\begin{eqnarray}
 (\rho T^{\m})_{2}& =& \frac{1}{G^{55}} 
   \hh^{\n\q}\hh^{\la\s}[2\frac{G^{5\m}}{G^{55}}G_{\la\n}G_{\s\q}\g^5
                         +4{G^{\m}}_{\la}\g_{\n}G_{\s\q}] = \\ \nonumber
                  &=&-\frac{2}{G^{55}} \nonumber
          [\frac{G^{5\m}}{G^{55}}{\rm tr}\,(\hh^2)\g^5 +2(\hh^2)^{\m\n}\g_{\nu}]\\
                  &=& U^{\m}_{2}.
\end{eqnarray}

At cubic order in ${\cal H}$, 
\begin{equation} \label{eq:3}
      (\rho T^{\m})_{3}=\rho_{1} T^{\m}_{2}+\rho_{2} T^{\m}_{1}. \label{rhoT3}
\end{equation}
Let the common factor to be
$(\tilde{\cal H}^{\al\be}\tilde{\cal H}^{\n\q}\tilde{\cal H}^{\la\s})$. 
Since,
\begin{equation}
\g^{5}  ({\g_{\n\q\la\s}}^{\m}-2\frac{G^{\m5}}{G^{55}}
           {\g_{\n\q\la\s}}^{5})=-{\g_{\n\q\la\s}}^{\m}\g^{5},
\end{equation}
we get
\begin{eqnarray*}
\rho_{1} T^{\m}_{2}& \sim& \frac{1}{4 G^{55}G_{5}}\bar{\g}
               \g_{\al\be}{\g_{\n\q\la\s}}^{\m} \gamma^5\\  
   &     \sim&   \frac{2}{G^{55}G_{5}}\bar{\g}
      (- \g_{\n\q\la} G_{\al\s}\delta^{\mu}_{\beta}
    + \frac{1}{2} {\g_{\n\q}}^{\m} G_{\al\s}G_{\be\la}
     - {\g_{\n\la}}^{\m}G_{\al\s}G_{\be\q})\g^{5}.
\end{eqnarray*}
The second term in eq.~(\ref{rhoT3}) can also be simplified:
\begin{eqnarray*}
\lefteqn{\q_{2} T^{\m}_{1} \sim - \frac{1}{4G^{55}G_{5}}\bar{\g}
               \g_{\n\q\la\s}(G^{\m5}\g_{\al\be}+
                                   2{G^{\m}}_{\be}\g_{\al}\g^{5})}\\\
        & &\sim  - \frac{2}{G^{55}G_{5}}\bar{\g}
   [ G^{\m5}(- \frac{1}{2} {\g_{\n\q}} G_{\al\la}G_{\be\s}
   + {\g_{\n\la}}G_{\al\q}G_{\be\s})
   -\g_{\n\q\la} G_{\al\s}\delta^{\mu}_{\beta}\g^{5}].
\end{eqnarray*}
Thus,
\begin{eqnarray}
(\rho T^{\m})_{3}&=& \frac{2}{G^{55} G_{5}}\bar{\g}
   [{\g_{\n\q}}^{\m}\g^{5} -G^{\m5}\g_{\n\q}]\nonumber
   [\frac{1}{2}{\hh}^{\n\q} {\rm tr}\,({\hh}^2) -({\hh}^3)^{\n\q}]\\
&=&  \frac{1}{G^{55}G_{5}}\nonumber
   \epsilon^{\al\be\n\q\m}\g_{\al\be}
   [\frac{1}{2}{\hh} {\rm tr}\,({\hh}^2) -{\hh}^3]_{\n\q}\\
&=&U^{\m}_{3}.
\end{eqnarray}

Finally, in the quartic order,
\begin{equation}
 \q_{2} T^{\m}_{2} = - \frac{G}{16 G_5^2} 
            \hh^{\m_{1}\m_{2}} \hh^{\m_{3}\m_{4}}
           \hh^{\n_{1}\n_{2}} \hh^{\n_{3}\n_{4}}  \nonumber
         \g_{\m_{1}\m_{2}\m_{3}\m_{4}}
     ( {\g_{\n_{1}\n_{2}\n_{3}\n_{4}} }^{\m} - 
     2 \frac{G^{\m5}}{G^{55}} {\g_{\n_{1}\n_{2}\n_{3}\n_{4}}}^{5}).
\end{equation}
The relevant contribution of $\g$'s in this case is 
\begin{eqnarray} 
\g_{\m_{1}\m_{2}\m_{3}\m_{4}}
         {\g_{\n_{1}\n_{2}\n_{3}\n_{4}}}^{\m} \sim \nonumber
         [8\g^{\m}(G_{\m_{1}\n_{1}}G_{\m_{2}\n_{2}}G_{\m_{3}\n_{3}} G_{\m_{4}\n_{4}} -
                   2G_{\m_{2}\n_{1}}G_{\m_{3}\n_{2}}G_{\m_{4}\n_{3}} G_{\m_{1}\n_{4}})\\
         -32\g_{\n_{1}}(\delta^{\mu}_{\mu_1}G_{\m_{2}\n_{2}}
G_{\m_{3}\n_{3}} G_{\m_{4}\n_{4}} -
                   2\delta^{\mu}_{\mu_2}G_{\m_{3}\n_{2}}G_{\m_{4}\n_{3}} G_{\m_{1}\n_{4}})]. \nonumber
\end{eqnarray}
It follows that
\begin{eqnarray}
 (\q T^{\m})_{4}&=&-\frac{G} {(G_5)^2}\nonumber
         \{ (\g^{\m}-2 \frac{G^{\m5}}{G^{55}}\g^5)
          [\frac{1}{2}({\rm tr}\,(\hh^2))^2 -({\rm tr}\,(\hh^4)] \\ \nonumber
      && - 4\g_{\n}[\frac{1}{2}{\rm tr}\,(\hh^2)\hh^2 -\hh^4]^{\m\n}\}\\
   &=&U^{\m}_{4}.
\end{eqnarray}
This completes the proof.

\end{document}